# Biological neurons act as generalization filters in reservoir computing

Takuma Sumi[a,b], Hideaki Yamamoto[a,*], Yuichi Katori[c,d], Satoshi Moriya[a], Tomohiro Konno[e], Shigeo Sato[a], Ayumi Hirano-Iwata[a,b,f]

[a] Research Institute of Electrical Communication (RIEC), Tohoku University, Sendai 980-8577, Japan.

[b] Graduate School of Biomedical Engineering, Tohoku University, Sendai 980-8577, Japan.

[c] Graduate School of System Information Science, Future University Hakodate, Hokkaido 041-8655, Japan.

[d] Institute of Industrial Science, The University of Tokyo, Tokyo 153-8505, Japan.

[e] Graduate School of Pharmaceutical Sciences, Tohoku University, Sendai 980-8578, Japan.

[f] Advanced Institute for Materials Research (WPI-AIMR), Tohoku University, Sendai 980-8577, Japan.

*To whom correspondence may be addressed: Hideaki Yamamoto

**Email:** hideaki.yamamoto.e3@tohoku.ac.jp

**Competing Interest Statement:** The authors declare no competing interest.

**This PDF file includes:**

    Main Text
    Figures 1 to 5
    Supplementary Figures S1 to S3

**Abstract**

Reservoir computing is a machine learning paradigm that transforms the transient dynamics of high-dimensional nonlinear systems for processing time-series data. Although the paradigm was initially proposed to model information processing in the mammalian cortex, it remains unclear how the non-random network architecture, such as the modular architecture, in the cortex integrates with the biophysics of living neurons to characterize the function of biological neuronal networks (BNNs). Here, we used optogenetics and fluorescent calcium imaging to record the multicellular responses of cultured BNNs and employed the reservoir computing framework to decode their computational capabilities. Micropatterned substrates were used to embed the modular architecture in the BNNs. We first show that modular BNNs can be used to classify static input patterns with a linear decoder and that the modularity of the BNNs positively correlates with the classification accuracy. We then used a timer task to verify that BNNs possess a short-term memory of ~1 s and finally show that this property can be exploited for spoken digit classification. Interestingly, BNN-based reservoirs allow transfer learning, wherein a network trained on one dataset can be used to classify separate datasets of the same category. Such classification was not possible when the input patterns were directly decoded by a linear decoder, suggesting that BNNs act as a generalization filter to improve reservoir computing performance. Our findings pave the way toward a mechanistic understanding of information processing within BNNs and, simultaneously, build future expectations toward the realization of physical reservoir computing systems based on BNNs.

**Main Text**

**Introduction**

The brain is a high-dimensional nonlinear system that exhibits intricate dynamics in response to sensory input and is responsible for essential functions, such as perception and motor control (1–6). These functions are physically implemented in a recurrent network of biological neurons, which are integrate-and-fire units realized by ion channel proteins inserted in a bilayer lipid membrane. Inherently, ion channels, cell membranes, and synaptic connections are all sources of biological noise, which add stochasticity to membrane potential (7) and signal transmission (8). Despite stochasticity in biological neurons, the population responses are relatively robust. For example, visual stimulation has been shown to stably recruit robust population activity (9, 10) or sequences of population activity (11, 12) in the mouse visual cortex. Consistent sequential activation of neuronal populations evoked by sensory stimuli has also been observed in the auditory and somatosensory cortex (13). Similarly, motor movement and decision-making have been shown to be associated with robust population dynamics in the motor cortex (14, 15) and posterior parietal cortex (16).

As a conceptual paradigm that links such transient population dynamics of high-dimensional nonlinear systems to information processing, 'reservoir computing' has been proposed in machine learning (17–19). The model generally consists of input, reservoir, and output layers, and its main feature is that only the weights between the reservoir and output layers are trained in a supervised manner, while the internal weights of the reservoir layer are fixed (19, 20). The tasks covered by reservoir computing include image classification, speech recognition, time-series prediction, and memory. Although reservoir computing was initially proposed to replace computationally demanding recurrent neural networks, it has been shown to have an affinity with many physical systems that exhibit complex dynamics (21, 22), including photonic circuits (23, 24), spintronic devices (25, 26), and soft/organic materials (27, 28). Cultured neuronal networks on uniform substrates, which exhibit a low-pass filtering property on input signals (29), have also been employed as reservoirs to demonstrate their capabilities in pattern classification (30, 31) and motor control (32).

Since the internal weights of the reservoir are fixed during training and inference, its network architecture critically affects the performance of a reservoir computing model.

Interestingly, multiple theoretical studies have recently shown that the implementation of bio-inspired connectivity, such as the modular structure (33, 34), small-worldness (35, 36), and scale-free degree distribution (36) that characterize the nervous systems of animals (37), can be advantageous in reservoir computing models of artificial neural networks. For example, Rodriguez et al. explored the optimal modularity of the reservoir layer in a memory task and showed that the highest short-term memory was achieved at a moderate modularity of approximately 0.2 (33). Kawai et al. adjusted the Watts-Strogatz rewiring probability in the reservoir layer and showed that small-world networks reduced the error of nonlinear time-series prediction (35). Furthermore, Suarez et al. showed that direct mapping of the human brain connectome to the reservoir maximize the memory capacity of a network and that any rewiring from the empirical network degrades the reservoir performance (38). Among the several canonical features of the nervous systems, the modular structure is especially important, as it is found both macroscopically at the level of cortical areas (39, 40) and microscopically at the level of neurons and synapses (41, 42). Modular networks are 'small-world' by their nature, balancing short path lengths and large clustering coefficients (43), and are also evolutionarily conserved (44). However, it remains unclear how such a network architecture integrates with the biophysics of living neurons, e.g., inherent noise and the filtering property, to characterize the dynamical function of biological neuronal networks (BNNs), due partially to the lack of an appropriate experimental platform.

In this study, we used rat cortical neurons cultured on micropatterned substrates combined with optogenetic tools to establish a simple *in vitro* model of BNNs to investigate how functional modularity affects the information processing capabilities of BNNs under the reservoir computing framework. We then investigated how BNNs transform input signals in a high-dimensional state space, focusing on generalization, or categorization, which is the ability of the brain to identify commonalities in different inputs (45). In dynamical terms, this is associated with consistency, wherein a subtle variation in the input signal is canceled to generate a similar dynamical trajectory (46). We show that functional modularity increases the separation of dynamical trajectories and improves the performance in pattern classification tasks and that BNNs act as generalization filters in such tasks to ease transfer learning. Generalization capability is inherent in the brains of living animals, while being a major challenge for machine learning systems (47). Our work paves the way toward a mechanistic understanding of information processing within BNNs and, simultaneously, builds future expectations toward the realization of physical reservoir computing systems based on BNNs.

## Results

**Reservoir computing with biological neurons.** Our reservoir computing system consists of an input layer, reservoir layer, and output layer. The core of the system, the reservoir layer, is a micropatterned biological neuronal network (mBNN) created by growing primary rat cortical neurons on an engineered substrate fabricated using microcontact printing (48) or a microfluidic device (49). Micropatterns mimicked the modular architecture of the nervous system of the animals and consisted of four squares (200 µm × 200 µm), wherein adhered neurons formed dense interconnections, connected by thin lines on which neurites grew to couple neighboring squares (Fig. 1*A*). Spontaneous and evoked activity of mBNN was measured by fluorescent calcium imaging using the calcium probe GCaMP6s. Input to the mBNN was delivered by irradiating patterned light to the neurons expressing red-shifted channelrhodopsin ChrimsonR (Fig. 1*B*). The output layer was implemented offline in a custom Python script, using a linear decoder.

First, we evaluated the classification performance of the mBNN reservoir using a three-class classification task on static spatial patterns (Fig. 1*C*). The output weights $W$ were trained by ridge regression, which generated a matrix that summed up the state $x(t)$ from the majority of the neurons comprising the reservoir (Fig. 1*D*). The output vector $y(t) = Wx(t)$ was then summed over time, and an element with the largest value was regarded as the estimated answer from the reservoir system. The accuracy of each mBNN reservoir was evaluated by dividing the number of correct estimates by the number of test trials.

The classification accuracy of the mBNN reservoirs after training was significantly higher than the value obtained from a label-shuffled null model (Fig. 1*E*). Fig. S1*A* summarizes the comparison of the accuracy using either the raw calcium fluorescence, the relative fluorescence unit (RFU), or their *z*-scored values as the source of the reservoir state, $x(t)$. Independent of the signal source, the mBNN reservoir could be successfully trained as a spatial pattern classifier (raw data, 71.9 ± 23.8%; *z*-scored raw data, 72.4 ± 26.6%; RFU, 74.8 ± 20.2%; *z*-scored RFU, 72.9 ± 25.3%; label-shuffled, 32.4 ± 13.0%; mean ± SD, $n$ = 21 samples).

Furthermore, the evaluation of the mean squared error (MSE) showed that the deviation of the output *y*(*t*) from the corresponding target signal was significantly smaller than that in a label-shuffled null model (Fig. 1*F*). The MSE significantly increased when the signal, either raw or RFU, was standardized by *z*-scoring (Fig. S1*B*) (raw data, 0.055 ± 0.022; *z*-scored raw data, 0.104 ± 0.020; RFU, 0.047± 0.016; *z*-scored RFU, 0.119 ± 0.018; mean ± SD, *n* = 21 samples). While the mean MSEs was statistically insignificant whether the raw fluorescence or RFU was used as *x*(*t*), the former was susceptible to baseline fluctuations. Thus, we used RFU as *x*(*t*) in all subsequent experiments.

**Modularity increases the accuracy of the spatial pattern classification.** Next, we investigated how non-random connectivity, especially modular organization, impacts the classification performance of mBNN reservoirs. To assess the functional connectivity of mBNNs, we evaluated the correlation coefficient between neurons *i* and *j*, $r_{ij}$, from the recordings of spontaneous activity and calculated the modularity *Q* of the correlation matrix [$r_{ij}$] in each network. Fig. 2*A* shows the correlation matrices of two representative mBNNs with high (*Q* = 0.13) and low (*Q* = 0.014) modularity. Notably, the functional connectivity (correlation matrix) of the highly modular network was characterized by the existence of subpopulations of neurons with strong coupling inside the four squares. Such substructures were absent in the functional connectivity of weakly modular networks, whose $r_{ij}$ was ~1 for almost all neuron pairs. Analysis of the relationship between functional modularity and computational performance revealed that classification accuracy was highly variable in networks with low modularity (<0.05), with a trend toward a positive correlation (Fig. 2*B*). In contrast, for networks with high modularity (>0.05), the accuracy was mostly independent of modularity, and the classification accuracy was consistently high (≥ 80%). The results suggest that functional modularity within mBNNs is advantageous for the classification of static spatial inputs with a linear readout.

To understand the mechanisms underlying the improved accuracy in highly modular networks, we analyzed the trajectories of the reservoir dynamics *x*(*t*) in response to each input (Fig. 2*C*). The similarity between the two trajectories was quantified by calculating their time-averaged Euclidean distances (see Materials and Methods) (Fig. 2*D*). Our analysis revealed that the mean distance between trajectories evoked by identical inputs of the same class ($D_{same}$) was 0.98 ± 0.59, while that for different classes ($D_{diff}$) was 1.1 ± 0.72 and was significantly larger (two-tailed *t*-test, *p* < 0.05; *n* = 21; Fig. 2*E*). We regarded the ratio of $D_{diff}$ to $D_{same}$ as the separability of the reservoir and investigated its dependence on the functional modularity. As summarized in Fig. 2*F*, modularity was significantly correlated with separability ($r$ = 0.74, $p$ < 0.001). These results indicate that the increased separability of reservoir dynamics underlies the improved classification performance of highly modular networks.

**Performance of mBNN reservoirs in speech recognition.** The strength of reservoir computing is its ability to process time-series data using the transient dynamics of the reservoir (18). In pattern recognition tasks, the reservoir must possess 'short-term memory' (50) to retain the information of the past input for a finite period. Such a reservoir property can be quantified using the timer task, a simple task that recalls the same signal as the input $u(t)$ with a delay $\tau$ (27, 51). Fig. 3 (inset) shows the target signals $\hat{y}_\tau = u(t - \tau)$ with $\tau$ = 0.5, 1, and 2 s, together with output signals $y_\tau$ from a representative mBNN reservoir trained to generate $\hat{y}_\tau$. As a general trend, the deviation of $y_\tau$ from $\hat{y}_\tau$ increases with $\tau$. The similarity between $y_\tau$ and $\hat{y}_\tau$ was quantified using the coefficient of determination $r^2$. The dependence of $r^2$ on $\tau$ for all networks ($n$ = 20) is summarized in Fig. 3. The $r^2$ value peaked at 50 ms, and then decreased with increasing $\tau$. The mean value of $r^2$ was higher than 0.1 within the range $\tau$ = [0 s, 1.2 s]. These results indicate that mBNNs have short-term memory on the order of seconds and suggest that they can effectively be used as a reservoir that processes time-series data with a timescale of ~1 s.

We thus designed a classification task for a human spoken digit dataset, TI-46, whose signal durations matched the short-term memory of the mBNN reservoir. To deliver spoken digit signals into the reservoir layer, the signal was first converted to a 17-channel cochleagram using Lyon's passive ear model. Illumination patterns were then designed by mapping each channel of the cochleagram to individual neurons expressing ChrimsonR and adjusting the spot size of the illumination according to its instantaneous spectral intensity (Fig. 4*A*). Spoken digit signals of 'zero' and 'one' were used in this study. The weight matrix of the output layer was then trained using ridge regression to generate a pulse signal corresponding to the input class, and the performance was evaluated following the same procedure as in the spatial pattern classification task.

Visualization of representative reservoir trajectories **x**(*t*) in the principal component (PC) subspace showed that mBNNs generated characteristic responses for the two input classes (Fig. 4*B*). The mean accuracy in classifying the two digits spoken by a female speaker was found to be 82.5 ± 21.9% (mean ± SD, *n* = 20) (Fig. 4*C*), which was significantly higher than the value obtained from label-shuffled null models. This result indicates that the mBNN reservoir can classify not only static spatial patterns but also complex spatiotemporal patterns, such as human spoken digits.

Although the majority of mBNNs showed a high classification accuracy of over 80%, we found that the accuracy exhibited a large network-to-network variation, ranging from 17% to 100%. Thus, to understand the mechanism that supports high accuracy reservoirs, we assessed the relationship between memory capacity ($= \sum_\tau r^2(\tau)$, where $r^2(\tau)$ is the coefficient of determination between $y_\tau$ and $\hat{y}_\tau$) and the accuracy of each mBNN. Interestingly, memory capacity did not correlate with accuracy (Fig. S2; *n* = 11), which we interpret to be caused by the sufficiently high memory capacity of all samples for classifying spoken digit signals. Thus, we analyzed the relationship between functional modularity *Q* and accuracy (Fig. 4*D*). Clearly, a large variability in accuracy was observed mostly in weakly modular networks (*Q* < 0.05), whereas the majority (= 5/6) of strongly modular networks (*Q* > 0.05) exhibited an accuracy of 100%. This result is consistent with that obtained from the spatial pattern classification task, suggesting that functional modularity is beneficial for both spatial and spatiotemporal pattern classifications.

Surprisingly, a comparison of the accuracy in the spatial pattern and spoken digit classification tasks revealed that the samples that exhibited high performance in one task did not necessarily perform well in the other task (Fig. 4*E*; *r* = 0.12, *n* = 11). Here, again, functional modularity plays an essential role. Fig. 4*F* shows the simultaneous probability of obtaining correct estimates in the two tasks as a function of modularity, showing that the simultaneous probability is positively correlated with *Q*. The results indicate that mBNN reservoirs with a high *Q* are stable classifiers independent of tasks.

**Generalization capability of mBNN reservoirs.** Finally, to test the generalization capability of the mBNN reservoir, we designed a generalized spoken digit classification task, wherein the reservoir was first trained to classify the 'zero' and 'one' inputs of a female or a male speaker and then tested its classification accuracy on 'zero' and 'one' inputs of a different gender (Fig. 5*A*). The input spectrum of the female/male speakers was distinct, and when the classifier was trained to regress directly from the photostimulation patterns illuminated on a silicon wafer (without the mBNN), the classification accuracy decreased to the chance level (50%) if the speaker was switched in the training and testing phases (Fig. 5*B*).

In contrast, when the signal was passed through the mBNN reservoir, the classification accuracy was only weakly affected by the switching of the speakers, and a mean accuracy of 64.1 ± 15.1% (*n* = 16) was retained (Fig. 5*C*). The value was 18.8% lower than that obtained without switching (82.8 ± 15.4%), but above the chance level, and significantly higher than the value obtained from the label-shuffled models. These results indicate that the classification is achieved non-randomly, and the signal transformation by the mBNN endows the reservoir with a generalization ability to distinguish spoken digits independent of the speakers.

To provide further evidence for the generalization ability of mBNN, we designed a gender classification task, wherein the reservoir was first trained to classify the spoken digit 'zero' of a male and a female speaker and then tested to classify the gender of the speaker on the spoken digit 'one' (Fig. 5*D*). Similar to the previous task, without the mBNN, an output layer trained on photostimulation images failed to classify gender when the spoken digits were switched (Fig. 5*E*). In contrast, the signal transformation by the mBNN increased the classification accuracy to 71.4%, which was well above the chance level. Although the accuracy was lower than that obtained without the switching of the digits, it was higher than that of the label-shuffled model (Fig. 5*F*). These results support the finding that the mBNN acts as a generalization filter for the spatiotemporal input patterns.

To understand which features of mBNNs enabled generalization, we investigated how each input signal was transformed after passing through the mBNN. The trajectories of photostimulation patterns $u(t)$ and reservoir layer $x(t)$ projected onto 3-dimensional PC subspaces are shown in Fig. 5*G*. The trajectories of $u(t)$ for the same digits spoken by the same speaker were constrained to a narrow tube, whereas those for different digits and speakers were separated. In contrast, the trajectories of $x(t)$ showed a large variability even for the same input classes but still formed widened tubes, which were separated for different classes. The mean Euclidean distance of the trajectories for the same class ($D_{same}$) was found to be 0.20 and 0.82 for $u(t)$ and $x(t)$, respectively. For different classes, the value ($D_{diff}$) was 1.76 and 0.98 for $u(t)$ and $x(t)$, respectively (Fig. 5*H*). This indicates that an mBNN transforms its input to reduce the difference between $D_{\text{diff}}$ and $D_{\text{same}}$ while retaining the magnitude relation of $D_{\text{diff}} > D_{\text{same}}$. Such a transformation enabled the generalized classification of signals of different genders or digits.

Finally, we quantified how the switching of digits and speakers affects the dynamics of the mBNN reservoir by evaluating the following three variables: $D_{\text{digits}}$, the mean distance between the trajectories $x(t)$ for the same speakers but different digits; $D_{\text{speaker}}$, that for the same digits but different speakers; and $D_{\text{both}}$, that for different speakers and digits. Distance analysis revealed that $D_{\text{both}}$ was significantly greater than both $D_{\text{digits}}$ and $D_{\text{speaker}}$ (Fig. 5*l*). This indicates that the separation between different classes was quantitatively retained in $x(t)$, exhibiting the largest separation when both digits and speakers were different. Therefore, the mBNN reduces the separation of trajectories between different classes in the high-dimensional state space while preserving the commonalities between different classes, enabling the distinction of different speakers and digits through a simple, linear decoder.

**Discussion**

In this study, we constructed mBNNs with modular topology on engineered substrates and demonstrated the effectiveness of the modular architecture and biological signal transduction in pattern classification tasks under the reservoir computing framework. Our experiments demonstrated that the accuracy of pattern classification was positively correlated with the functional modularity of BNNs. Conventionally, a reservoir computing system based on artificial neural networks takes advantage of a sparsely connected random network as the reservoir layer to exploit its high-dimensional dynamics (19, 52). BNNs grown on uniform substrates similarly form random networks, although some metric dependence exists due to the finite length of the axons. However, the connectivity between neurons is usually dense, and spontaneous and evoked activity patterns of BNNs in uniform culture are typically governed by the so-called network bursts, i.e., repetitive firing of neurons that occur coherently across a large population of neurons in a network (53–55). Inducing modular topology in BNNs has been shown to break the excessive coherence in cultured BNNs and increase the variety of intrinsic activity patterns (48, 49, 56), which is probably the underlying mechanism behind the positive correlation between functional modularity and classification accuracy in mBNNs.

We investigated the short-term memory of the mBNN reservoir using a timer task and showed that mBNNs retain a short-term memory of ~1 s. This value is mostly in agreement with the report by Dranias et al., which showed that the short-term memory in uniformly cultured BNNs in a stimulus-specific memory task was ~1 s (57). Kubota et al. reported a memory capacity of 20 ms in uniformly cultured BNNs (58), but the difference of that value from that in Dranias et al. is most likely caused by the task design. Despite the differences in the reports, characteristic timescales of tens of milliseconds to several seconds, which are comparable to physical reservoirs based on soft materials (27), organic molecules (28), or water in a bucket (59), have been consistently observed in BNNs. This time scale matches that of many real-world signals, such as visual and auditory signals, making BNNs a suitable substrate for physical reservoirs in practical applications.

The mBNN was successfully demonstrated to be capable of classifying not only static spatial patterns but also spatiotemporal signals, such as human spoken digits, with a linear decoder. This contrasts with a previous pioneering study in which a nonlinear classifier (support vector machine) was used to classify spatiotemporal data such as jitter spike train and random music delivered to a uniform BNN (31). While it remains to be investigated whether modularity in BNNs increases performance in other tasks in reservoir computing, such as motor control (32, 60), signal generation (20), and time-series prediction tasks (23), previous computational studies on artificial and spiking neural networks bearing modular structures (33, 35, 38) strongly imply that modularity in mBNN can improve the performance of these tasks as well.

Most importantly, our experiments demonstrated that mBNNs act as a generalization filter by reducing the trajectory distance between common classes and increasing the distance between separate classes, enabling classification by readout layers pretrained on different genders or digits. Such 'categorical learning' was impossible when the input signal was directly decoded without an mBNN. In machine learning, data augmentation techniques such as noise addition, scaling, and rotation have been used to improve the generalization ability by reducing overfitting and expanding the decision boundary of the trained model (61). Although data augmentation has mostly been used for image classification in artificial neural networks, more recently, the technique has been reported to improve the generalization capability in time-series classification (62, 63). We hypothesize that the inherent noise in biological neurons intrinsically performs data augmentation to improve the generalization capability of the BNN reservoir.

Reservoir computing is an elegant framework that enables the establishment of a link between the high-dimensional dynamics of complex systems and their computational functions. By applying the concept of physical reservoirs, we constructively investigated how BNNs transform input signals and how their input responses can be utilized for computing. The mBNN reservoir could pave the way toward providing a platform to investigate structure-function relationships in BNNs, develop brain-like physical reservoirs, and understand the mechanisms of diseases with abnormalities in network modularity and population dynamics, such as autism (64) and epilepsy (65).

**Materials and Methods**

**Fabrication of micropatterned biological neuronal networks**. mBNNs were constructed by culturing primary neurons on microfabricated coverslips prepared by microcontact printing. A glass coverslip (C018001, Matsunami) was first sonicated in 100% ethanol and Milli-Q water, dried under $N_2$ flow, and subsequently treated in air plasma for 60 s (PM-100, Yamato). The surface of the coverslip was then rendered non-permissive to cell adhesion by coating with a 0.2% solution of poly(2-methacryloyloxyethyl phosphorylcholine-co-3-methacryloxypropyl triethoxysilane) (66) in ethanol for 10 s, drying in an ethanol environment for 20 min, baking in an oven at 70 °C for 4 h, and further drying in a vacuum chamber overnight. The coated coverslip was rinsed in ethanol to remove excess physisorbed molecules and sterilize the sample.

The protein ink used for microcontact printing was a mixture of extracellular matrix gel (E1270, Sigma-Aldrich; 1:100 dilution) and poly-D-lysine (50 µg ml$^{-1}$; P0899, Sigma-Aldrich) and was patterned using a polydimethylsiloxane (PDMS) stamp. The fabrication of the PDMS stamp and stamping procedure have been detailed previously (53, 67). Each micropattern comprised four small squares (200 µm × 200 µm) connected by a thin line of approximately 5–10 µm. After stamping the protein ink, four pieces of PDMS (2 mm × 2 mm × 0.5 mm) were placed at the periphery of the coverslip as spacers. After drying the patterned protein ink overnight on a clean bench, the coverslip was finally immersed in neuronal plating medium [minimum essential medium (MEM; 11095-080, Gibco) + 5% fetal bovine serum + 0.6% D-glucose]. In some experiments, PDMS microfluidic films bearing identical micropattern structures were fabricated and attached to a poly-D-lysine coated coverslip, as described previously (49), and were used to replace microcontact printing for neuronal patterning.

**Cell culture and transfection.** All procedures involving animal and gene transfection experiments were approved by the Tohoku University Center for Laboratory Animal Research, Tohoku University (approval number: 2020AmA-001), and Tohoku University Center for Gene Research (approval number: 2019AmLMO-001). Primary neurons were obtained from the cortices of embryonic day 18 pups, plated on the microfabricated coverslip, and co-cultured with astrocyte feeder cells in 5 mL of N2 medium [MEM + N2 supplement + ovalbumin (0.5 mg ml$^{-1}$) + 10 mM HEPES]. At 4 days in vitro (DIV), half of the medium was removed, and a red-shifted channel rhodopsin ChrimsonR (Addgene viral prep #59171-AAV9) (68) using adeno-associated virus vectors was added at a concentration of 2.0 µL. At 5 DIV, the medium that was removed at 4 DIV was added back, together with 1 mL of Neurobasal medium [Neurobasal (21103-049, Gibco) + 2% B-27 supplement (17504-044, Gibco) + 1% GlutaMAX-I (35050-061, Gibco)]. Half of the medium was replaced with the Neurobasal medium at 8 DIV.

**Optogenetic stimulation.** Stimulation of cultured neuronal networks was performed using optogenetics by activating ChrimsonR with patterned LED light (Thorlabs Solis 623C; nominal wavelength, 623 nm). Patterned light was created using a digital mirror device (DMD; Mightex Polygon400-G) coupled to an LED via a liquid light guide. The DMD was mounted on an inverted microscope (Olympus IX-83), and the patterned light was reflected onto the sample stage using a short-pass dichroic mirror with an edge wavelength of 556 nm (Semrock FF556-SDi01).

For the spatial pattern classification task, three distinct positions with identical areas (167×200 µm$^2$) were selected from the cultured neuronal network and irradiated for 100 ms at 10 s intervals. The photostimulation patterns were created and uploaded to Polygon400-G using PolyScan2 software (Mightex), and the switching of the patterns was controlled by externally generated transistor–transistor logic signals. The three patterns were delivered 30 times in random order in each session.

For the spoken digit classification task, photostimulation patterns were designed using a custom MATLAB script. First, the spoken digit data of a female and a male saying 'zero' and 'one' were selected from the TI-46 dataset and converted to a cochleagram by Lyon cochlear filter using the following the filter setting: decimation factor = 250, quality factor = 2, filter stepping factor = 0.25, and number of frequency channels = 17. To convert the cochleagram to photostimulation patterns, each of the frequency channels in the cochleagram was mapped one-to-one to individual neurons expressing ChrimsonR, and the spectrum intensity of the cochleagram was normalized by the maximum value in the entire cochleagram and converted to circular patterns such that the value of the normalized cochleagram $v \in [0,1]$ corresponds to a diameter of 25×$v$ μm. Finally, black-and-white bitmap files containing the photostimulation patterns were generated and imported into PolyScan2 software. The photostimulation patterns corresponding to 'zero' and 'one' were alternately irradiated every 15 s for a total of 20 times in each session.

**Calcium Imaging.** At 10–12 DIV, the coverslips with mBNN were rinsed in HEPES-buffered saline (HBS) containing 128 mM NaCl, 4 mM KCl, 1 mM $CaCl_2$, 1 mM $MgCl_2$, 10 mM D-glucose, 10 mM HEPES, and 45 mM sucrose, and subsequently loaded with 2 μM fluorescent calcium indicator Cal-520 AM (AAT Bioquest) in HBS for 30 min in an incubator. The coverslip was then transferred to a glass-bottom dish (3960-035, Iwaki) filled with fresh HBS. The dynamics of the calcium indicator fluorescence in response to optogenetic stimulation were imaged using an inverted microscope (IX83, Olympus) equipped with a 20× objective lens (numerical aperture, 0.75), a light-emitting diode light source (Lambda HPX, Sutter Instrument), a scientific complementary metal-oxide semiconductor camera (Zyla 4.2P, Andor), and an incubation chamber (Tokai Hit). All the recordings were performed at 37 °C. Up to three mBNNs were selected from a coverslip, and fluorescence imaging was performed at 20 frames/s using the Solis software (Andor). In some experiments, spontaneous activity was recorded for 6 min prior to recording stimulation responses. All values in the analyzed data are presented as the mean ± SD.

**Reservoir Computing.** To use the dynamics of an mBNN as the reservoir state, $N$ (= 60) neurons (15 neurons per module) were manually selected from a single network, and their somas were defined as regions of interest (ROIs). The relative fluorescence unit (RFU) $\Delta F/F$ of cell $i$ at time $t$ was calculated as $[F_i(t) - F_0]/F_0$, where $F_i(t)$ is the mean intensity within ROI *i* at time *t* and $F_0$ is the baseline fluorescence of the ROI. Δ*F*/*F* was then used to construct the network dynamics ***x***(*t*) (= [*x_i*(*t*)]) as *x_i*(*t*) = [*F_i*(*t*) − $F_0$]/$F_0$.

The reservoir computing model consisted of an input layer, a reservoir layer, and a readout layer. Input ***u***(*t*) was delivered optogenetically, as described above. The reservoir layer was an mBNN that generated an *N*-dimensional signal ***x***(*t*), which was determined by the intrinsically generated spontaneous activity and by its response to ***u***(*t*). The readout layer was a three- or two-dimensional linear decoder for the spatial pattern and spoken digit classification tasks, respectively. The readout state at time *t*, ***y***(*t*), is calculated as:

$$\boldsymbol{y}(t) = \boldsymbol{W}\boldsymbol{x}(t),$$

where ***W*** denotes the output weight matrix. ***W*** was trained using ridge regression as:

$$\boldsymbol{W} = \boldsymbol{X}^\mathrm{T}\widehat{\boldsymbol{Y}}(\boldsymbol{X}\boldsymbol{X}^T + \lambda \boldsymbol{I})^{-1},$$

where ***X*** = [***x***(1), …, ***x***(*t*), …] is the reservoir state matrix, $\widehat{Y}$ is the target signal matrix, *λ* (= 1) is the regularization coefficient, and ***I*** is the identity matrix. An element of the corresponding class in the target signal matrix was set to one for a duration of 2.5 s and zero otherwise. The number of data used for the training and test phases were 20 and 10, respectively, for the spatial pattern classification task and 14 and 6, respectively, for the spoken digit classification task. During the test phase, the estimated answer was obtained from ***y***(*t*) as $\mathrm{argmax}_i \left[\sum_{t=t_k}^{t_k+\Delta t} y_i(t)\right]$, where *i* is an element in the output layer, $t_k$ is the onset of the *k*th input, and Δ*t* is the duration of the target (set to 2.5 s). Accuracy was evaluated as the fraction of correct estimates during the test phase.

In the timer task, the network was stimulated with one of the three spatial patterns lasting for 100 ms, and the network response ***x***(*t*) was regressed to produce a delayed target signal using ridge regression. The delayed target signal was a square wave identical to the input signal, whose onset was delayed by *τ* s against the onset of stimulation. The short-term memory of the neuronal network was evaluated using the coefficient of determination $r^2(\tau)$ as (50):

$$r^2(\tau) = \frac{\mathrm{Cov}^2[u(t-\tau), y_\tau(t)]}{\mathrm{Var}[u(t)] \cdot \mathrm{Var}[y_\tau(t)]},$$

where Cov is the covariance, Var the variance, $u(t)$ the input, and $y_\tau(t)$ the output when the target signal is delayed by $\tau$.

**Trajectory Analysis.** To quantify the similarity/dissimilarity between the trajectories, the distance between the two trajectories $x(t)$ and $y(t)$, $D_{xy}$, was calculated as follows:

$$D_{xy} = \langle d_{xy}(t) \rangle_t = \langle \|x(t) - y(t)\|_2 \rangle_t$$

where $d_{xy}(t)$ is the distance between trajectories $x$ and $y$ at time $t$, $\|\cdot\|_2$ and $\langle \cdot \rangle_t$ denote the L2 norm and temporal average, respectively. The mean distance of the trajectories in response to the same (different) input classes was defined as $D_{\text{same}}$ ($D_{\text{diff}}$) and evaluated as follows:

$$D_{\text{same}} = \langle D_{xy} \rangle \, (c_x = c_y)$$

$$D_{\text{diff}} = \langle D_{xy} \rangle \, (c_x \neq c_y)$$

where $c_i$ is the input class of trajectory $i$ ($x \neq y$), and $\langle \cdot \rangle$ is the mean across the corresponding trajectory pairs.

**Network Analysis.** For the analysis of functional connectivity within mBNNs, the firing rate of neuron $i$, $f_i$, was inferred from the relative fluorescence intensity $\Delta F/F$ using the CASCADE algorithm with a convolutional neural network trained with the Global_EXC dataset (sampling rate = 20 Hz, smoothing window = 100 ms) (69). The Pearson correlation coefficient between neurons $i$ and $j$, $r_{ij}$, was calculated as follows:

$$r_{ij} = \frac{\text{Cov}[f_i(t), f_j(t)]}{\sqrt{\text{Var}[f_i(t)] \cdot \text{Var}[f_j(t)]}}$$

The functional modularity $Q$ of the correlation matrix was then quantified using Newman's method (70) as follows:

$$Q = \frac{1}{2M} \sum_{ij} \left( r_{ij} - \frac{k_i k_j}{2M} \right) \delta(m_i, m_j)$$

where *M* is the sum of all weighted edges (= $\frac{1}{2}\sum_{ij} r_{ij}$), $k_i$ is the sum of the weighted edges attached to neuron *i* (= $\sum_j r_{ij}$), δ is the Kronecker delta, and $m_i$ is the module containing neuron *i*.

**Data, Materials, and Software Availability.** Datasets generated and analyzed during the current study, along with all study-specific reagents and codes, are available from the corresponding author upon request. All other study data are included in the article and/or supporting information.

**Acknowledgments**

We thank Kei Wakimura and Taiki Takemuro for setting up the microscope. We acknowledge MEXT Grant-in-Aid for Transformative Research Areas (B) 'Multicellular Neurobiocomputing' (21H05163, 21H05164), JSPS KAKENHI (18H03325, 20H02194, 20K20550, 21J21766, 22H03657, 22K19821), JST PRESTO (JMPJPR18MB), JST CREST (JPMJCR19K3), and Tohoku University RIEC Cooperative Research Project Program for financial support.

**Figures and Tables**

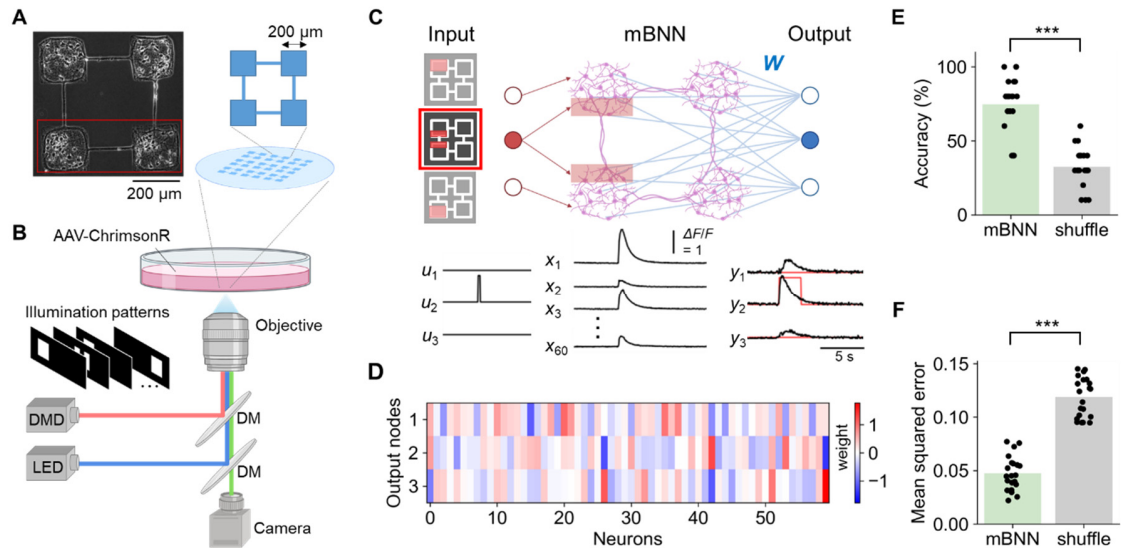

**Fig. 1.** Reservoir computing based on mBNN. (*A*) Phase-contrast micrograph of a micropatterned biological neuronal network. Photostimulation was delivered to neurons located in the lower half of the network outlined with a red rectangle. (*B*) Schematic representation of the experimental setup. DMD, digital mirror device; LED, light-emitting diode; DM, dichroic mirror. (*C*) Overview of the reservoir computing system. (*D*) A representative weight matrix of the output layer. (*E*) Classification accuracy and (*F*) MSE of the mBNN reservoir computed using the relative fluorescence unit as $x(t)$. 'Shuffle' denotes the values obtained using label-shuffled datasets. Filled circles represent a single mBNN, and bar heights show the mean. ***$p < 0.001$ (two-sided unpaired *t*-test).

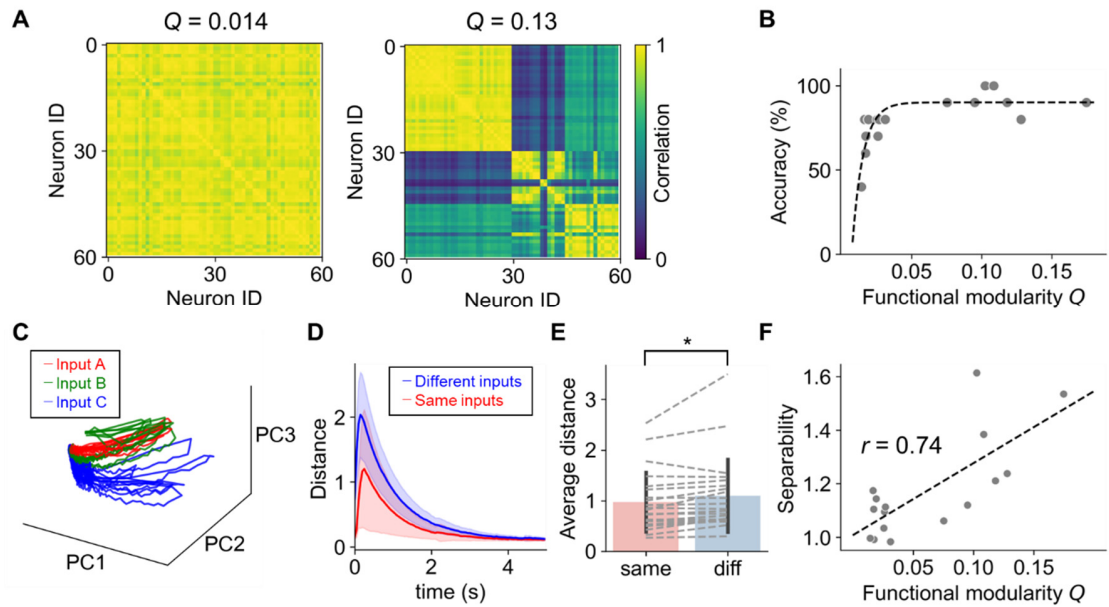

**Fig. 2.** Structure-function relationships in mBNN reservoirs. (*A*) Representative correlation matrices of two mBNNs with weak (left) and strong (right) modularity. Correlations between two neurons are evaluated using the recordings of spontaneous activity. (*B*) Classification accuracy in the spatial pattern classification task as a function of functional modularity. Weakly modular networks show large variability in accuracy, whereas strongly modular networks stably show high accuracy. A single exponential curve fit is also shown to aid visualization (dashed line). (C) Trajectories of reservoir dynamics $x(t)$ visualized in a 3-dimensional principal component subspace. (*D*) Instantaneous distance between two trajectories upon receiving the same or different inputs. Mean value at each time are plotted with shades indicating the SD. (*E*) Time-averaged distance between trajectories for the same and different inputs. Bar heights show the mean, error bars show the SD, and dotted gray lines show individual mBNN. *$p < 0.05$ (two-sided paired *t*-test). (*F*) Separability of the trajectories ($D_{diff}/D_{same}$) as a function of functional modularity.

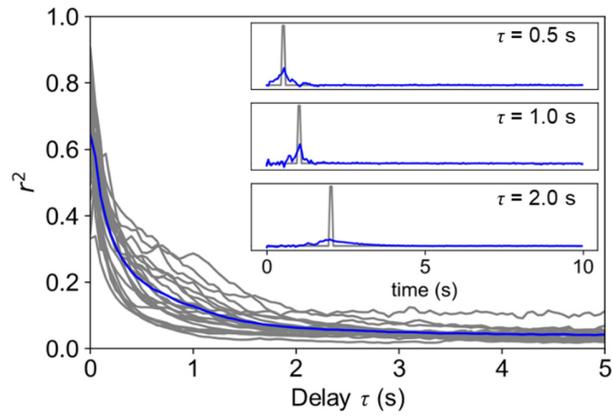

**Fig. 3.** Memory capacity of mBNNs. Coefficient of determination plotted as a function of time delay $\tau$. Gray curves represent a single mBNN, and a blue curve is the mean. Inset shows the training data (gray) and representative reservoir outputs (blue) at designated time delays. Deviation of the reservoir output from the training data increases with the time delay.

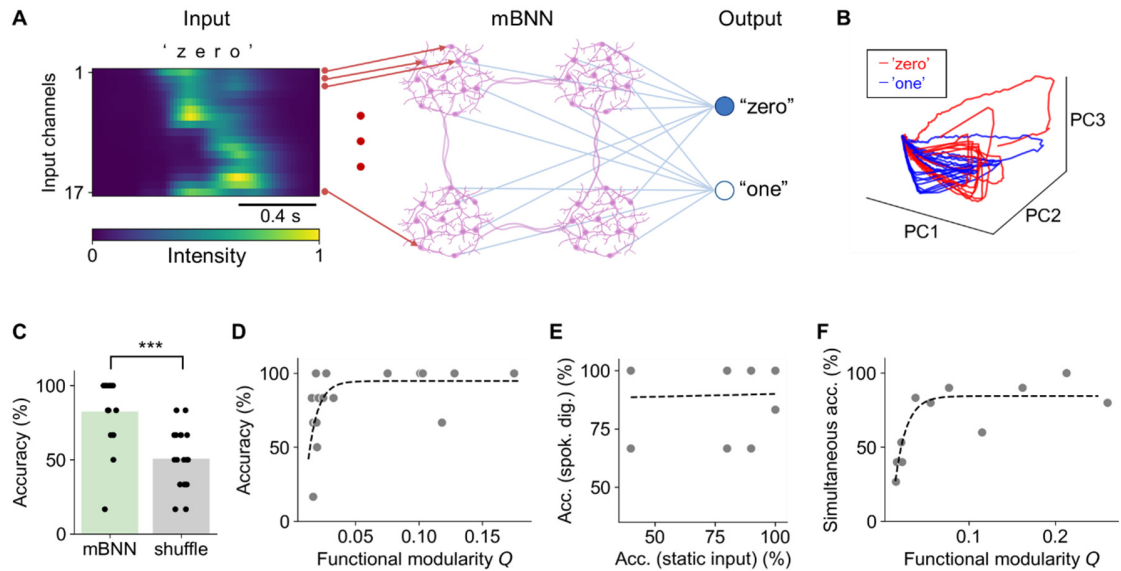

**Fig. 4.** Classification of the time-series data. (*A*) Overview of the spoken digit classification task. Audio waveforms are first converted to a 17-channel cochleagram, which are then mapped onto the neurons in mBNNs by assigning one neuron for one frequency band. The input signal is then optogenetically delivered to mBNNs, and the evoked activity is then decoded in the output layer. (*B*) Trajectories of reservoir dynamics $x(t)$ visualized in a 3-dimensional principal component subspace. (*C*) Classification accuracy of the mBNN reservoir. 'Shuffle' denotes the values obtained using label-shuffled datasets. Filled circles represent a single mBNN, and bar heights show the mean. ***$p < 0.001$ (two-sided unpaired *t*-test). (*D*) Classification accuracy in the spoken digit classification task as a function of functional modularity. As was for the spatial pattern classification task, weakly modular networks show large variability in accuracy, whereas strongly modular networks stably show high accuracy. A single exponential curve fit is shown to aid visualization (dashed line). (*E*) Accuracy in the spoken digit classification task does not correlate with the accuracy in the spatial pattern classification task. A linear fit is shown to aid visualization (dashed line). (*F*) Simultaneous accuracy, i.e., a product of the accuracies in the spoken digit and spatial pattern classification tasks, as a function of modularity. Note that the functional modularity calculated from evoked activity is used in the abscissa due to the limited number of samples with the recordings of spontaneous activity. The functional modularity calculated from the spontaneous and evoked activities were, however, strongly correlated (Fig. S3; $r = 0.80$ and $0.73$). A single exponential curve fit is shown to aid visualization (dashed line).

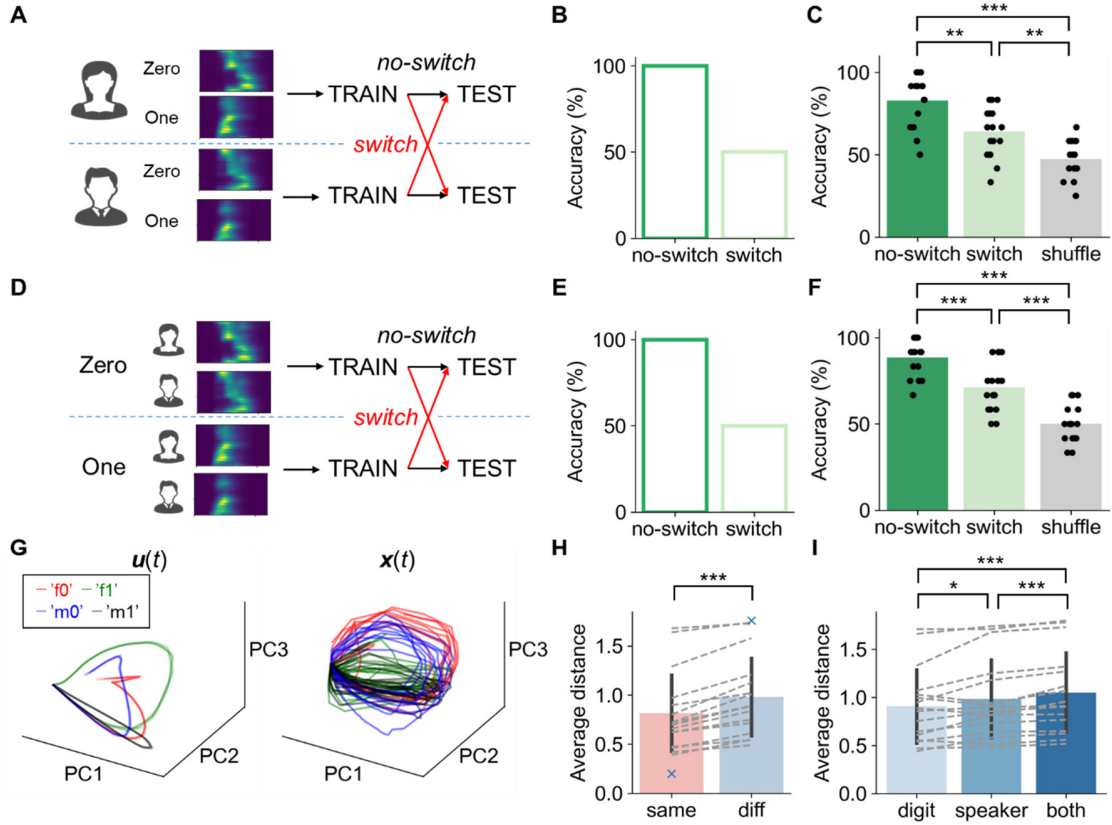

**Fig. 5.** Generalization ability of mBNN reservoirs. (A) Schematic illustration of the generalized spoken digit classification task. The output weight $W$ is first trained to classify the two digits in one gender, and the accuracy of classifying the digits spoken by a different gender is evaluated during the test phase. (B) The classification accuracy when the photostimulation patterns are directly decoded by a linear decoder. (C) The classification accuracy by the mBNN reservoir. The accuracy is highest when the same inputs are used in the training and testing phases ('no switch'). However, mBNN still retains an accuracy above chance level ('shuffled') even when the gender of the speaker is switched in the training and testing phases ('switch'). **$p < 0.01$; ***$p < 0.001$ (two-sided unpaired $t$-test). (D–F) Same as in (A–C) except for the gender classification task. (G) Trajectories of the photostimulation pattern $u(t)$ (left) and reservoir dynamics $x(t)$ (right) visualized in a 3-dimensional principal component subspace. m (male) and f (female) denote the gender, and 0 and 1 denote the digit. (H) Time-averaged distance between the trajectories $x(t)$ for the same and different inputs. Blue crosses indicate the distance calculated from the photostimulation pattern $u(t)$. Bar heights show the mean, and error bars show the SD. ***$p < 0.001$ (two-sided paired $t$-test). (I) Time-averaged distance between the trajectories $x(t)$ of different digits, different speakers, and different digits and speakers. *$p < 0.05$; ***$p < 0.001$ (two-sided paired $t$-test).

**Supplementary Figures**

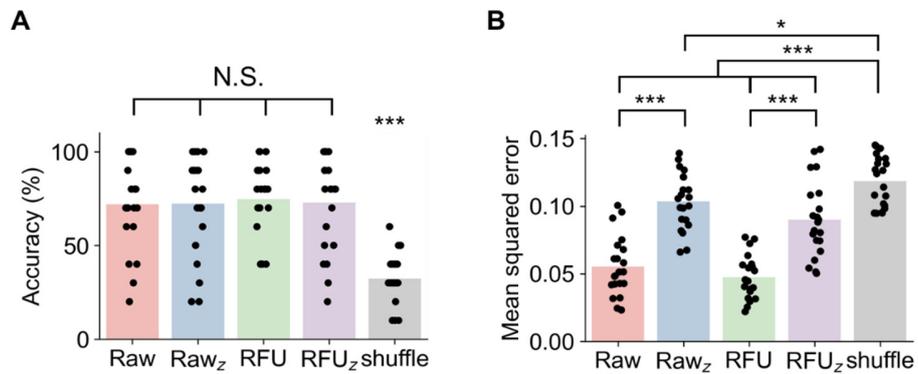

**Fig. S1.** (*A*) Classification accuracy and (*B*) mean squared error (MSE) of the mBNN reservoir computed using different signals as $x(t)$: Raw, raw fluorescence intensity; Raw$_z$, $z$-scored raw intensity; RFU, relative fluorescence unit; and RFU$_z$, $z$-scored RFU. 'shuffle' denotes the values obtained using label-shuffled datasets with the RFU as $x(t)$. The accuracy and MSE calculated from the label-shuffled datasets are significantly lower and higher, respectively, than all other data. Filled circles represent a single mBNN, and bar heights show the mean. *$p$ < 0.05; ***$p$ < 0.001; N.S., no significance (two-sided unpaired *t*-test).

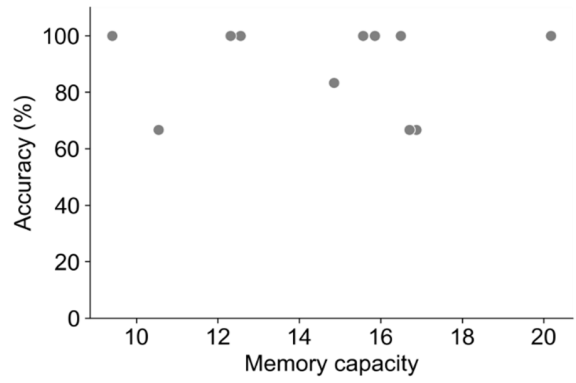

**Fig. S2.** The classification accuracy of the spoken digit classification task is independent of the memory capacity of the mBNNs.

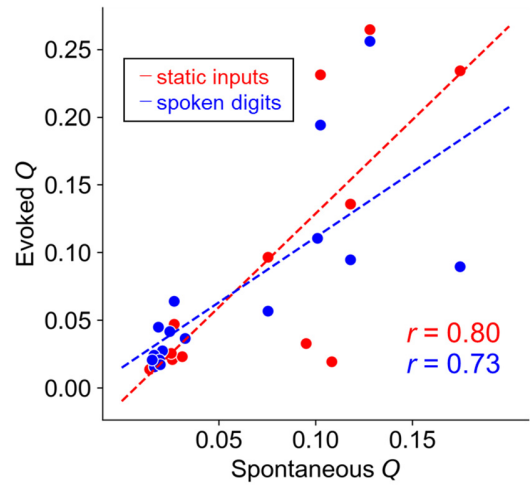

**Fig. S3.** Functional modularity $Q$ calculated from the evoked activity strongly correlates with the value of $Q$ calculated from the spontaneous activity recording in both the static pattern classification task (red) and the spoken digit classification task (blue).